\documentclass[prb,twocolumn,superscriptaddress,float,aps]{revtex4-2}
\usepackage{bm,color,amsmath,amssymb,mathrsfs,latexsym,graphicx,psfrag,tabularx}

\usepackage[hidelinks,colorlinks,linkcolor=blue,
citecolor=blue,urlcolor=blue]{hyperref}

\newcommand{\beq}{\begin{equation}}
\newcommand{\eneq}{\end{equation}}

\usepackage{etoolbox}
\makeatletter
\patchcmd{\@maketitle}{\@author}{\@author\show\@thanks}{}{}
\makeatother

\usepackage{dsfont}

\begin{document}

\title{Phonon circular birefringence and polarization-filter in Magnetic Topological Insulators}

\author{Abhinava Chatterjee}
\affiliation{%
   Department of Physics, The Pennsylvania State University, University Park, Pennsylvania 16802, USA
}
\author{Chao-Xing Liu}
\email{cxl56@psu.edu}
\affiliation{%
   Department of Physics, The Pennsylvania State University, University Park, Pennsylvania 16802, USA
}
\affiliation{Center for Theory of Emergent Quantum Matter, The Pennsylvania State University, University Park, Pennsylvania 16802, USA}

\begin{abstract} 
The surface phonon Hall viscosity (PHV) -an acoustic analog of axion electrodynamics-emerges from the strain response of magnetic topological insulators and gives rise to novel acoustic phenomena.In this work, we propose a previously unexplored effect: a phonon polarization–filter mechanism induced by the surface PHV, which generates an interface phonon mode with its frequency below the bulk mode frequency. This interface mode possesses a specific circular polarization and therefore acts as a polarization filter, confining only phonons with the matching polarization at the interface. Magnetic topological insulators can thus selectively transmit one type of circularly polarized phonon mode, enabling the manipulation of phonon polarization and angular momentum. In addition, we further develop a generalized scattering framework to study the effect of an injected acoustic wave from a trivial insulator to a magnetic topological insulator with  both normal and oblique incidence, and discuss the phenomena of surface acoustic Faraday rotation and longitudinal-transverse mode conversion. Our results establish surface Hall viscosity as a powerful mechanism for engineering axial phonon states and open new avenues for topological phononic devices based on phonon angular momentum. 
\end{abstract}

\date{\today}


\maketitle

\section{Introduction}

Phonons, quantized collective lattice vibrations, serve as one of the fundamental carriers of mechanical energy and information in solids. Beyond their frequency and momentum, phonons possess internal degrees of freedom such as spin or angular momentum \cite{vonsovskii1961spin2,levine1962note,auld1973acoustic,ren2022elastic} encoded in their polarization \cite{bommel1960excitation,anastassakis1972morphic1,anastassakis1972morphic2,anastassakis1972morphic3,schaack1975magnetic,dohm1975magnetoelastic,thalmeier1975rare,balkanski1976proceedings,schaack1976observation}. Recent advances have highlighted the potential for controlling phonon polarization \cite{zhang2014angular,hamada2018phonon,nakane2018angular,rebane1983faraday,lin1985study,mclellan1988angular,bermudez2008chirality,kagan2008anomalous,jones1973asymmetric,zhang2014angular,chen2018chiral,hamada2018phonon,chen2019chiral,chen2021propagating,ren2021phonon,saparov2022lattice,zhang2023gate,wang2024chiral,juraschek2025chiral,shabala2025axial,wang2025ab,wang2025alteraxial} and chirality \cite{pine1969linear,pine1971raman,yin2021chiral,zhang2023weyl,choi2022chiral,oishi2024selective,zhang2015chiral,zhu2018observation,ishito2023chiral,zhang2025measurement,che2025magnetic,yang2025inherent,ueda2023chiral,ueda2025chiral,romao2024phonon,fava2025phonon,zhang2024understanding,zeng2025photo,luo2023large,ishito2023truly,luo2023large,davies2024phononic,basini2024terahertz}, opening new avenues for several phononic functionalities \cite{vangessel2018review,maldovan2013sound,li2012colloquium,balandin2012phononics,liu2020topological}, such as adaptive phonon routing \cite{huang2024parity,hatanaka2024valley,patel2018single,ren2022topological}, polarization-based signal encoding \cite{wu2025magnetic}, and improved control of energy flow and information in acoustic and thermal devices \cite{nomura2022review}.  Phonon birefringence \cite{luthi1965ferroacoustic,kluge1965drehung,kluge1966akustische,portigal1968acoustical,pine1971linear,joffrin1970mise,bialas1982circular,vlasov1986magnetic,kouderis2022acoustically} — the polarization-dependent splitting of propagation velocities — provides a route to controlling phonon angular momentum through functionalities such as waveplates  \cite{wei2024elastic} and polarization converters \cite{marunin2022polarization}. In particular, phonon circular birefringence corresponds to the splitting of the propagation velocities between right and left circularly polarized phonon modes. Existing approaches to induce phonon birefringence typically rely on external fields \cite{sonntag2021electrical}, magneto-elastic coupling \cite{luthi1979magnetoelasticity,volluet1979optical,low2014magneto,rao1941magneto,muller2024chiral,skibin2014features,ozhogin1988anharmonicity,mirsaev1998nonlinear}, or microstructuring in metamaterials \cite{lee2024perfect,psarobas2014birefringent}. 

Magnetic topological insulators (TIs) \cite{tokura2019magnetic,liu2023magnetic,bernevig2022progress} host a rich variety of polarization-dependent acoustic phonon phenomena due to the their surface phonon Hall viscosity (PHV) \cite{barkeshli2012dissipationless,shapourian2015viscoelastic,avron1995viscosity,avron1998odd,chatterjee2026surface}, rooted in the bulk topological Nieh-Yan term \cite{hughes2011torsional,nieh1982identity,nieh1982quantized,chandia1997topological,nieh2007torsional}. For transverse acoustic waves propagating parallel to the external magnetic field or intrinsic magnetization, the surface PHV gives rise to circular phonon birefringence, manifesting as acoustic Faraday rotation \cite{boyd1966attenuation,boiteux1971acoustical,wang1971acoustic,thalmeier1978faraday,tucker1980acoustic,tucker1980theory,lee1999discovery,lee2000acoustic,halperin2000acoustic,frenzel2019ultrasound,tuegel2017hall,dominguez1995interaction,dominguez1996interaction,sytcheva2010acoustic,sytcheva2010magneto,thalmeier2009paramagnetic,tokman2011acoustic,tuegel2017hall} that exhibits rotation of their polarization vector across the surface of magnetic TIs \cite{shapourian2015viscoelastic}. A related effect, longitudinal-transverse mode conversion, occurs when an oblique incident longitudinal acoustic wave at the surface of magnetic TIs develops an additional transverse mode component due to the same surface PHV \cite{shapourian2015viscoelastic}. Both the acoustic Faraday rotation and the longitudinal-transverse mode conversion occur at the surface of magnetic TI materials and are induced by the surface PHV.

\begin{figure*}
\includegraphics[width=\textwidth]{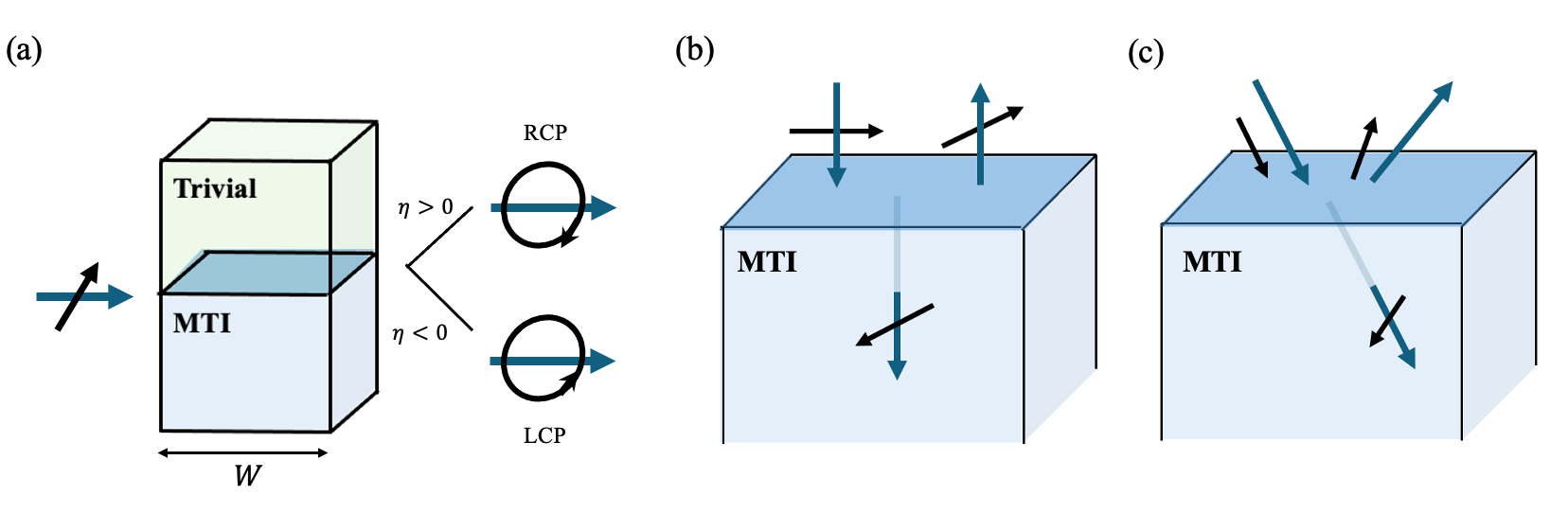}
 \caption{ (a)  The schematic of the proposed phonon polarization-filter based on the interface of magnetic TI structure. Injecting a linearly polarized wave leads to the detection of either a RCP or LCP wave near the interface, depending on the interface magnetization. (b) Schematic of the surface scattering phenomena -   acoustic Faraday rotation and the (c) longitudinal-transverse mode conversion. The blue and black arrows depict the propagation and polarization directions, respectively. 
 }
 \label{fig:Schematic}
 \end{figure*}

In this work, we propose a previously unknown interface-specific acoustic phonon effect: a polarization-filter mechanism for acoustic waves propagating parallel to the interface between a magnetic TI and a trivial insulator. This effect arises from the surface PHV, which gives rise to an interface-localized mode carrying a definite circular polarization. Thus far, the proposed phonon polarization filters have either separated phonon the right and left circular polarizations by coupling to photons with specific circular polarizations \cite{kim2025chiral,pine1969linear,pine1971linear,yin2021chiral,ishito2023truly,ishito2023chiral,zhang2023weyl,choi2022chiral} or by separating enantiomers \cite{oishi2024selective}. In contrast, we show that chiral phonon states with selective circular polarizations can be generated without photon-driven or crystal chirality-induced excitations, but due to interface confinement in magnetic TIs originating from the surface PHV, as shown in Fig. \ref{fig:Schematic}(a). By leveraging the bulk topology encoded in the surface PHV, the circular polarization of the interface phonon mode can be tuned by the material's internal magnetic order \cite{shapourian2015viscoelastic,chatterjee2026surface}. We propose a heterostructure to realize this phonon polarization-filter, as presented in Fig. \ref{fig:Schematic}(a) of a trivial TI and a magnetic TI, such as magnetically doped TI sandwiches (e.g. Cr doped TI/pure TI/V doped TI sandwiches) \cite{yu2010quantized,chang2013experimental,zhang2012tailoring,morimoto2015,xiao2018realization,mogi2017magnetic,mogi2017tailoring,zhuo2023axion,jiang2020concurrence,kou2015magnetic} and MnBi$_2$Te$_4$ \cite{deng2020quantum,liu2020robust,otrokov2019prediction,zhang2019topological,li2019intrinsic,gong2019experimental,liu2020robust}.

To provide a unified theoretical framework, we develop a generalized scattering theory of acoustic waves at the surface of realistic magnetic topological insulators by extending previous treatments of acoustic Faraday rotation and mode conversion \cite{shapourian2015viscoelastic} and incorporating all symmetry-allowed, anisotropic PHV coefficients determined by the $D_{3d}$ point group of magnetic TIs. These known effects naturally emerge as limiting cases of our theory, while phonon polarization-filter effect appears as a qualitatively new consequence unique to the surface PHV response with no counterpart in the axion electrodynamics response of magnetic TIs. The remainder of the paper is organized as follows. In Sec. \ref{sec:Phonon dynamics}, we review the theoretical framework for the effective phonon equation of motion arising due to the surface PHV, which was previously developed in Ref. \cite{chatterjee2026surface}. In Sec. \ref{sec:Phonon birefringence}, we present the phonon polarization-filter mechanism arising due to the interface PHV in magnetic TI materials (Fig.\ref{fig:Schematic}(a)). In Sec. \ref{sec:Faraday}, we generalize the acoustic Faraday rotation presented in Ref. \cite{shapourian2015viscoelastic} to incorporate all the symmetry-allowed terms for a realistic magnetic TI material (Fig.\ref{fig:Schematic}(b)). In Sec. \ref{sec:Conversion}, we generalize the mode conversion mechanism presented in Ref. \cite{shapourian2015viscoelastic} including anisotropic surface PHV coefficients (Fig.\ref{fig:Schematic}(c)). In Sec. \ref{sec:Conclusion}, we conclude with a discussion on possible experimental detections.

\begin{figure*}
\includegraphics[width=\textwidth]{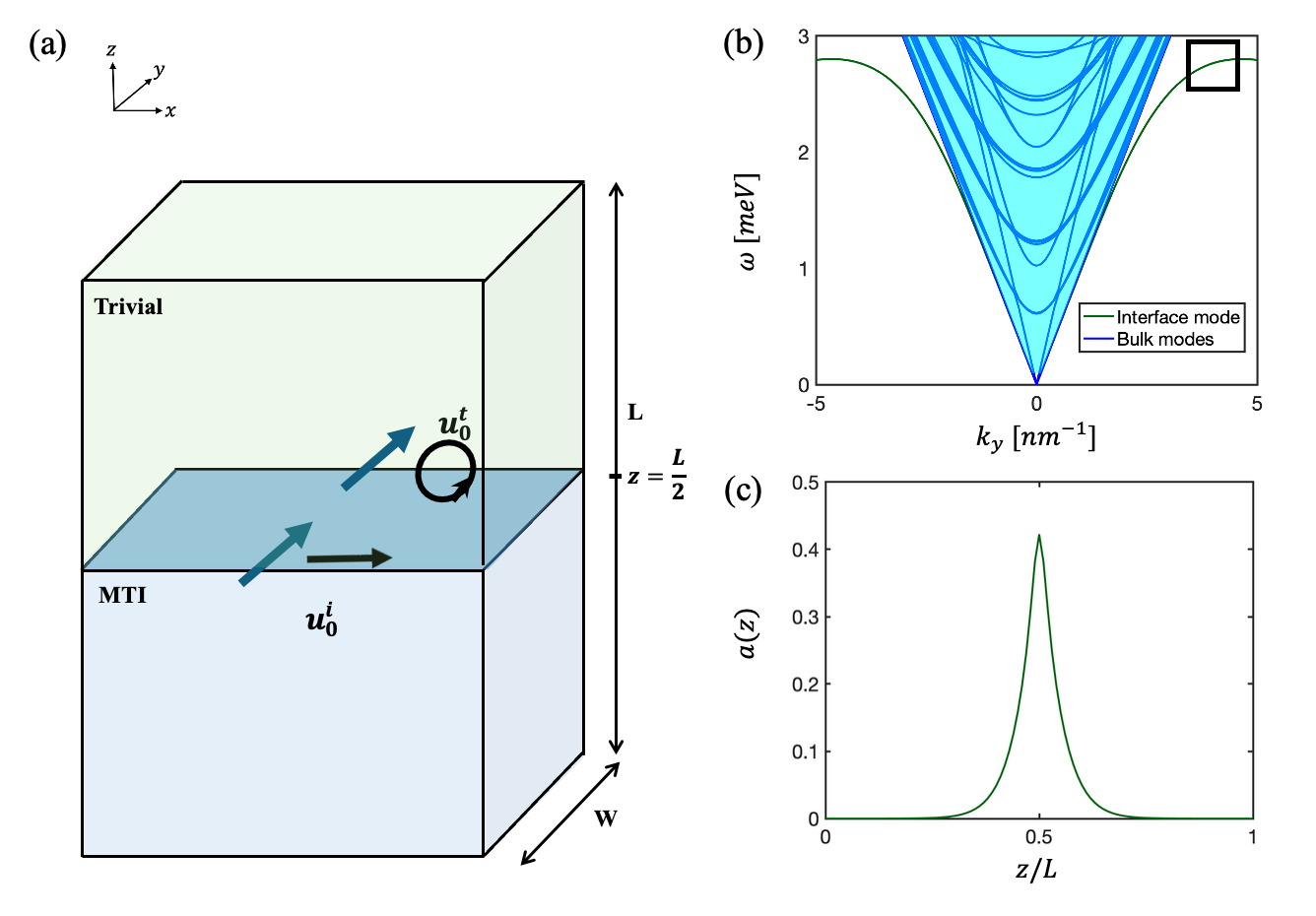}
 \caption{(a) The setup of the phonon polarization-filter, where the injected wave at $y=0$ is linearly polarized along the $x$-direction and the transmitted wave detected at $y=W$ is either RCP or LCP. (b) The phonon frequency in the isotropic approximation. The interface phonon mode is depicted in green and the bulk modes are in blue. (c) The $z$-profile of the interface mode at $k_y = 4$ nm$^{-1}$. We use $c_t = 1500, c_l = 2500$ in units of m/s and $\eta_1 = 0.16 \eta_0, \eta_2 = 0.11 \eta_0, \eta_3 = 0.07 \eta_0$ where $\eta_0$ is in units of $\frac{1}{\rho_0} \frac{1}{\hbar} \frac{1}{(\mu m)^2}$ 
 }
 \label{fig:CM1}
 \end{figure*}

\section{Phonon Dynamics} \label{sec:Phonon dynamics}
We consider the phonon dynamics at the interface of a magnetic TI and a trivial TI with the same elastic modulii. The effective phonon equation of motion has previously been derived in Ref.\cite{chatterjee2026surface},
\begin{equation}\label{eq:EOMgeneral}
    \rho \ddot{u_i}  = \partial_j \left( \lambda_{ijkl} u_{kl} - 2 \eta_{ijkl} \frac{\partial \Phi}{\partial z} \dot{u}_{kl} \right), 
\end{equation} 
where $\rho$ is the mass density, $u_{ij}$ is the strain tensor defined as $u_{ij} = (\partial_i u_j + \partial_j u_i)/2$, $\lambda_{ijkl}$'s are the bulk elastic modulii, $\eta_{ijkl}$'s are the coefficients of PHV\cite{barkeshli2012dissipationless, shapourian2015viscoelastic,guo2021extrinsic,chen2020enhanced,cortijo2015elastic,Heidari2019optical,ye2020phonon,huang2021electron,zhang2021phonon,chatterjee2026surface}  and $\Phi$ is the phase of the complex Dirac mass. The bulk elastic modulii satisfy $\lambda_{ijkl} = \lambda_{jikl} = \lambda_{ijlk} = \lambda_{klij}$. For magnetic TI materials with $D_{3d}$ point group symmetry, the bulk elastic modulii are given by \cite{landau2012theory}
\begin{eqnarray}\label{eq:lambda}
   && a+b \equiv \lambda_{xxxx} = \lambda_{yyyy}  \nonumber \\
   && a-b \equiv \lambda_{xxyy} \nonumber \\
   && b \equiv \lambda_{xyxy} \nonumber \\
   && c \equiv \lambda_{xxzz} = \lambda_{yyzz} \nonumber \\
   && d \equiv \lambda_{xzxz} = \lambda_{yzyz} \nonumber \\
   && g \equiv \lambda_{xxyz} = -\lambda_{yyyz} = \lambda_{xyxz}.
\end{eqnarray}
The values of bulk elastic modulii for magnetic TI sandwich structures are given by the parameter values in Bi$_2$Te$_3$ for $a,b,c,d,f,g$, according to Ref.\cite{jenkins1972elastic}. For MnBi$_2$Te$_4$, some of the elastic modulii have been presented in Refs. \cite{chai2024thermoelectric,bartram2022ultrafast}, but since a complete list is missing, we will use the elastic modulii of Bi$_2$Te$_3$ \cite{jenkins1972elastic} due to their similar crystal structure. The PHV satisfies the symmetry relation upon permutation of indices, 
\begin{eqnarray}\label{eq:symmetry_eta}
    \eta_{ijmn} = \eta_{jimn} = \eta_{ijnm} = -\eta_{mnij}.
\end{eqnarray}
 Due to Eq.(\ref{eq:symmetry_eta}) and the $D_{3d}$ symmetry, there are three independent PHV coefficients, given by
\begin{align} \label{eq:eta symmetry}
    & \eta_1 \equiv \eta_{xxxy} = -\eta_{yyxy} = -\eta_{xyxx} = \eta_{xyyy}  \nonumber \\
    & \eta_2 \equiv \eta_{xzyz} = -\eta_{yzxz} \nonumber \\
    & \eta_3 \equiv \frac{\eta_{xyyz}}{2} = -\eta_{xxxz} = \eta_{yyxz} = -\frac{\eta_{yzxy}}{2} = \eta_{xzxx} = -\eta_{xzyy}. 
\end{align}
In Ref.\cite{shapourian2015viscoelastic}, an isotropic surface PHV was considered, with $\eta_1 = \eta_2$ and $\eta_3 = 0$, and its effects on acoustic Faraday rotation and mode conversion were investigated. In contrast, we generalize the theoretical formalism and examine the effects of all the three independent surface PHV coefficients $\eta_{1,2,3}$. We first consider periodic boundary conditions along $x$ and $y$ directions and an interface at the origin along the $z$ direction, so $\Phi(z) = \pi \Theta(-z)$ where $\Theta(z)$ is the Heaviside step function. Here we consider an interface between magnetic TI material and a trivial insulator with the same symmetry group and bulk elastic moduli such that the existence of localized phonon modes purely stems from the change in $\Phi$. We will solve the scattering problem for the equation of motion (\ref{eq:EOMgeneral}) for the above interface configuration. 

We first consider a solution with the plane wave in the in-plane (xy) direction and spatial variation in the z-direction, so we choose the ansatz 
\begin{eqnarray}\label{eq_SM:ansatz_elasticwave}
\textbf{u}(\textbf{r},t) = \textbf{f}(z)e^{i \textbf{k}_{\parallel}\cdot\textbf{r}_{\parallel}}
\end{eqnarray}
with $\textbf{r}_{\parallel} = (x,y)$ and $\textbf{k}_{\parallel} = (k_x,k_y)$. Substituting the ansatz Eq.(\ref{eq_SM:ansatz_elasticwave}) into Eq.(\ref{eq:EOMgeneral}), we find the eigen equation
\begin{eqnarray}\label{eq_SM:eigen_elasticwave}
    H_{\text{ph}}(\omega) \textbf{f} = \omega^2 I \textbf{f}
\end{eqnarray}
where $I$ is an identity matrix and
\begin{equation}\label{eq:H2iw}
    H_{\text{ph}}(\omega) = H_0 + 2 i \omega  H_{\text{PHV}}
\end{equation}
with
\begin{widetext}
\begin{equation}
    H_0  = \begin{pmatrix} 
     (a+b) k_x^2 + b k_y^2 - d \partial_z^2 -2i g k_y \partial_z  & a k_x k_y -i 2g k_x \partial_z  &  2g k_x k_y -i (c+d) k_x \partial_z \\
    a k_x k_y - i 2g k_x \partial_z  & b k_x^2 +  (a+b) k_y ^2 - d \partial_z^2 + 2g i k_y \partial_z & g (k_x^2 - k_y^2) -i (c+d) k_y \partial_z  \\
     2g k_x k_y - i (c+d) k_x \partial_z & g (k_x^2 - k_y^2) - i (c+d) k_y \partial_z & d (k_x^2 + k_y^2)  - f \partial_z^2 
    \end{pmatrix},\label{eq:H0}
\end{equation}
and
\begin{equation}
    H_{\text{PHV}}  =  \begin{pmatrix} 
     0  &  \frac{\partial \Phi}{\partial z}  \left( \eta_1 (k_x^2 + k_y^2) - \eta_2 \partial_z^2 \right) & -\frac{\partial \Phi}{\partial z}  \left(  \eta_2 (i k_y) \partial_z + \eta_3 (k_x^2- k_y^2) \right)\\ 
       -\frac{\partial \Phi}{\partial z}  \left( \eta_1 (k_x^2 + k_y^2) - \eta_2 \partial_z^2 \right) & 0 & \frac{\partial \Phi}{\partial z} \left(  \eta_2 (i k_x) \partial_z + 2 \eta_3 k_x k_y \right) \\
     \frac{\partial \Phi}{\partial z}  \left(  \eta_2 (i k_y) \partial_z + \eta_3 (k_x^2-k_y^2) \right)  & -\frac{\partial \Phi}{\partial z}  \left(  \eta_2 (i k_x) \partial_z + 2\eta_3 k_x k_y \right)  & 0
    \end{pmatrix}. \label{eq:HNY}
\end{equation}
\end{widetext}

The eigen-problem of Eq.(\ref{eq_SM:eigen_elasticwave}) can be solved numerically\cite{chatterjee2026surface}, and the resulting eigen-energies and eigen-wavefunctions are utilized to construct the solutions of the scattering problems described below.

\begin{figure*}
\includegraphics[width=\textwidth]{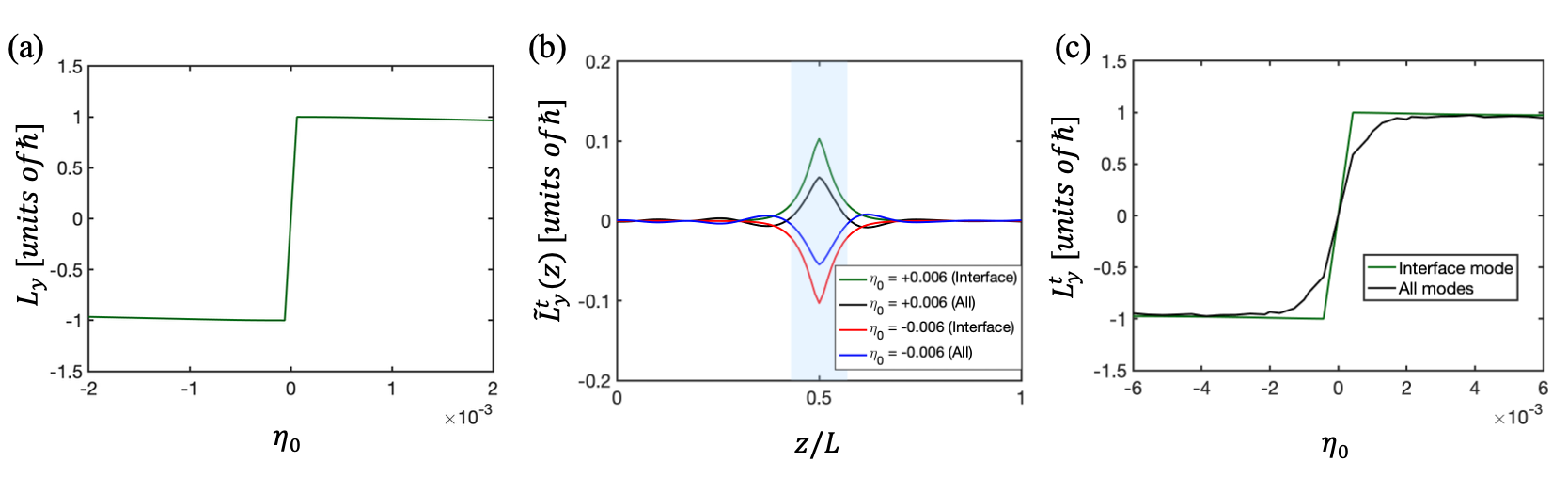}
 \caption{ (a) The phonon angular momentum $L_y$ in the $xz$-plane for the interface eigen-mode as a function of the surface PHV coefficient $\eta_0$. (b) The phonon angular momentum $\Tilde{L}_y^t(z)$ as a function of $z$ for the interface mode with $\eta_0 > 0$ (in green) and $\eta_0 < 0$ (in red). The $\Tilde{L}_y^t(z)$ due to a sum over all the contributing modes for $\eta_0 > 0$ (in black) and $\eta_0<0$ (in blue). Here, we use an injected wave frequency $\omega = 2.5$ meV.  (c) The phonon angular momentum $L_y^t$ in the $xz$-plane for the transmitted mode near the interface as a function of the surface PHV coefficient $\eta_0$. We separate the contributions from the interface mode (in green) and all the propagating modes (in black). We use $c_t = 1500, c_l = 2500$ in units of m/s and $\eta_1 = 0.16 \eta_0, \eta_2 = 0.11 \eta_0, \eta_3 = 0.07 \eta_0$ where $\eta_0$ is in units of $\frac{1}{\rho_0} \frac{1}{\hbar} \frac{1}{(\mu m)^2}$. 
}
 \label{fig:CM2}
 \end{figure*}

\section{Interface phonon polarization filter}\label{sec:Phonon birefringence}
In this section, we theoretically propose a phonon polarization-filter mechanism based on an interface phonon eigen-mode originating from the PHV. The interface mode is either right circularly (RCP) or left circularly polarized (LCP) depending on the intrinsic magnetic order at the interface. We compute this numerically for our slab model (along z) with the interface PHV terms in Eq.(\ref{eq:EOMgeneral}). In this section, we consider the isotropic approximation given by 
\begin{equation} \label{eq:isoparameters}
    a =  c_l^2 - c_t^2, b = c_t^2, c = (c_l^2 - 2 c_t^2), d = c_t^2, f = c_l^2, g=0,
\end{equation}
where $c_{t,l}$ are the transverse and longitudinal velocities. As illustrated in Fig.\ref{fig:CM1}(a), when the elastic wave is traveling along $y$-direction, Eqs.(\ref{eq:H0},\ref{eq:HNY}) at $k_x = 0, k_y \neq 0$ can be simplified as
\begin{widetext}
\begin{equation}\label{eq_SM:isoFaradH0}
    H_0  = \begin{pmatrix} 
      c_t^2 k_y^2 - c_t^2 \partial_z^2  & 0  &  0 \\
    0  & c_l^2 k_y ^2 - c_t^2 \partial_z^2  & -i (c_l^2 - c_t^2) k_y \partial_z  \\
     0 & - i (c_l^2- c_t^2) k_y \partial_z & c_t^2 k_y^2  - c_l^2 \partial_z^2 
    \end{pmatrix},
\end{equation}
and
\begin{equation} \label{eq:isoFilterPHV}
    H_{\text{PHV}}  =   \begin{pmatrix} 
     0  &  \frac{\partial \Phi}{\partial z}  \left( \eta_1 k_y^2 - \eta_2 \partial_z^2 \right) & -\frac{\partial \Phi}{\partial z}  \left(  \eta_2 (i k_y) \partial_z - \eta_3 k_y^2 \right)\\ 
       -\frac{\partial \Phi}{\partial z}  \left( \eta_1 k_y^2 - \eta_2 \partial_z^2 \right) & 0 & 0 \\
     \frac{\partial \Phi}{\partial z}  \left(  \eta_2 (i k_y) \partial_z - \eta_3 k_y^2 \right)  & 0  & 0
    \end{pmatrix}.
\end{equation}
\end{widetext}
The phonon dispersion is shown in Fig.\ref{fig:CM1}(b) where the interface phonon mode and the bulk phonon modes are represented by green and blue lines, respectively. The interface mode has a lower frequency than the bulk modes and thus is expected to be localized at the interface. In the slab model with N sites along the $z$-direction, the interface mode, denoted as ${\bf u}_{0}$ is of the form
 \begin{align}
    &{\bf u}_{0} = \begin{pmatrix}
        a(z_1) \\ a(z_2) \\ . \\ . \\ a(z_N)
    \end{pmatrix} \otimes \frac{1}{\sqrt{2}}\begin{pmatrix}
        i \\ 0 \\ -1 
    \end{pmatrix} e^{i k_y y - i \omega t},\label{eq:Nonzeroetamodes}
\end{align}
where the $a(z)$ profile is shown in Fig.\ref{fig:CM1}(c), in which the interface phonon mode is localized near the interface $z/L \sim 0.5$. The total out-of-plane phonon angular momentum  \cite{vonsovskii1961spin2,zhang2014angular,hamada2018phonon} is given by $L_y = 2 \hbar \sum_m \text{Im} \left( u^*_z(z_m) u_x(z_m) \right) $, where the sum $m$ is over the N sites. From Eq.(\ref{eq:Nonzeroetamodes}), the interface mode carries finite out-of-plane angular momentum, $L_y = +\hbar$ corresponding to right circular polarization (RCP) in the $xz$-plane. The polarization of the interface phonon mode is shown in Fig.  \ref{fig:CM2}(a) as a function of the surface PHV $\eta_0$, where we treat $\eta_0$ as a tuning parameter by rewriting the surface PHV coefficients as $\eta_1 = 0.11 \eta_0, \eta_2 = 0.11 \eta_0, \eta_3 = 0.07 \eta_0$. Since the sign of the surface PHV is determined by the sign of the magnetization at the interface \cite{shapourian2015viscoelastic,chatterjee2026surface}, the polarization of the interface phonon modes is flipped when the magnetization is flipped, as shwon in Fig.\ref{fig:CM2}(a), where the polarization of the interface mode flips from RCP ($L_y = +\hbar$) to LCP ($L_y = -\hbar$) when $\eta_0$ goes from positive to negative. At zero surface PHV, $\eta_0$, the polarization vanishes ($L_y=0$). 

Next, we consider injecting a transverse elastic wave at $y=0$ propagating in the positive $y$-direction with linear polarization in the x direction and a Gaussian $z$-profile localized around the interface, which can be described by the function
\begin{equation}
    {\bf f}^i = \begin{pmatrix}
        g(z_1) \\ g(z_2) \\ . \\ . \\ g(z_N)
    \end{pmatrix} \otimes \begin{pmatrix}
        1 \\ 0 \\ 0
    \end{pmatrix}, \quad g(z_m) \sim e^{-\frac{\left( z_m - L/2 \right)^2}{2 \sigma^2}}.
\end{equation}
We follow Ref.\cite{chang1973surface} and treat the injected elastic wave as a source term in the equation of motion 
\begin{equation} \label{eq:Eqnsource}
    \omega^2 {\bf u} = H^{N}_{\text{ph}}(\omega,-i \partial_y) {\bf u} + \delta(y)  {\bf f}^i,
\end{equation}
where both ${\bf u, f}^i$ are $3N$ column vectors in the slab model with $N$ sites along the $z$-direction and $H^{N}_{\text{ph}}(\omega,-i \partial_y)$ is a $3N \times 3N$ matrix obtained from Eqs.(\ref{eq_SM:isoFaradH0},\ref{eq:isoFilterPHV}) by taking $k_y \rightarrow - i \partial_y$ and considering a slab configuration with N sites along the $z$-direction. In Eqs.(\ref{eq_SM:isoFaradH0},\ref{eq:isoFilterPHV}), we have set $k_x=0$ enforcing uniformity along the $x$-direction. Under this condition, the source term in Eq.(\ref{eq:Eqnsource}) reduces to a line source positioned at $y=0$. We define the Green's function of Eq.(\ref{eq:Eqnsource}) as 
\begin{equation} \label{eq:Greensfunction}
    \left( \omega^2 - H^N_{\text{ph}} \right) G(y) = \delta(y) .
\end{equation}
By Fourier transform to the momentum space, we have
\begin{equation}
    G(y) = \int \frac{dk_y}{2 \pi} e^{i k_y y} \Tilde{G}(k_y),
\end{equation}
where the (retarded) Green's function in momentum space $\Tilde{G}(k_y)$ is given by
\begin{equation}
    \Tilde{G}(k_y) = \sum_n \frac{{\bf f}_n(k_y)  {\bf f}^\dagger_n (k_y) }{ \left( \omega + i \epsilon \right) ^2 - \omega_n^2(k_y) },
\end{equation}
where $\epsilon$ is a small positive number ($\epsilon>0$), and ${\bf f}_n(k_y)$ and $\omega_n$ are the n$^{\text{th}}$ eigenstate and eigenvalue of $H^N_{\text{ph}}$, respectively. We choose $n=0$ for the interface state and $n=1, ..., 3N-1$ for other bulk states. The transmitted elastic wave at $y>0$ is determined by 
\begin{align}
    {\bf u}^t(y)  &= \int dy' G(y,y') \delta(y')  {\bf f}^i  \nonumber \\
    &= \sum_n \int \frac{dk_y}{2 \pi} e^{i k_y y} \frac{ \left( {\bf f}^\dagger_n (k_y)  {\bf f}^i \right)}{\left( \omega + i \epsilon \right)^2 - \omega_n^2(k_y) } {\bf f}_n(k_y)  . \label{eq:Psiy1}
\end{align}
We only consider $\omega>0$ and the poles of Eq.(\ref{eq:Psiy1}) are determined by 
\begin{equation}
    \omega = \omega_n(k_y),
\end{equation}
which occur at momenta $k_y = k_n$ for $n=0, 1,2,..., N_k-1$. The number of the solutions, $N_k$, satisfies $N_k \leq 3N$, since several high-energy phonon modes may have frequencies always above the injected wave frequency $\omega$ for all $k_y$. In this case, the corresponding solutions for $k_y$ are complex and can be written as $k_{n>N_{k}} = k^R_n + i \kappa_n$. Physically, they correspond to evanescent elastic waves that decay away from $y=0$. Consequently, their contribution to  ${\bf u}^t(y)$ cannot be detected at a large distance, $y = W \gg 1/\kappa_n$. Eq.(\ref{eq:Psiy1}) boils down to 
\begin{equation}
   {\bf u}^t(y) = -\frac{1}{2}  \sum_n  \int \frac{dk_y}{2 \pi} e^{i k_y y} \frac{  \left( {\bf f}^\dagger_n (k_y) {\bf f}^i \right)}{\omega \Big[ v_n(k_y) (k_y - k_n) - i \epsilon \Big]} {\bf f}_n(k_y)  , \label{eq:Psiy2}
\end{equation}
where $v_n(k_y) = d\omega_n(k_y)/d k_y$ is the phonon velocity of the n$^{\text{th}}$ mode. Eq.(\ref{eq:Psiy2}) has simple poles at $k_y = k_n + i\epsilon/v_n$ and using standard complex integral techniques, the phonon mode at $y>0$ becomes
\begin{equation} \label{eq:PsiyFinal}
 {\bf u}^t(y)  = - \frac{i}{2} \sum_n e^{i k_n y} \frac{ \left( {\bf f}^\dagger_n (k_n)  {\bf f}^i \right)}{\omega v_n(k_n) } {\bf f}_n(k_n) .
\end{equation}

\begin{figure*}
\includegraphics[width=\textwidth]{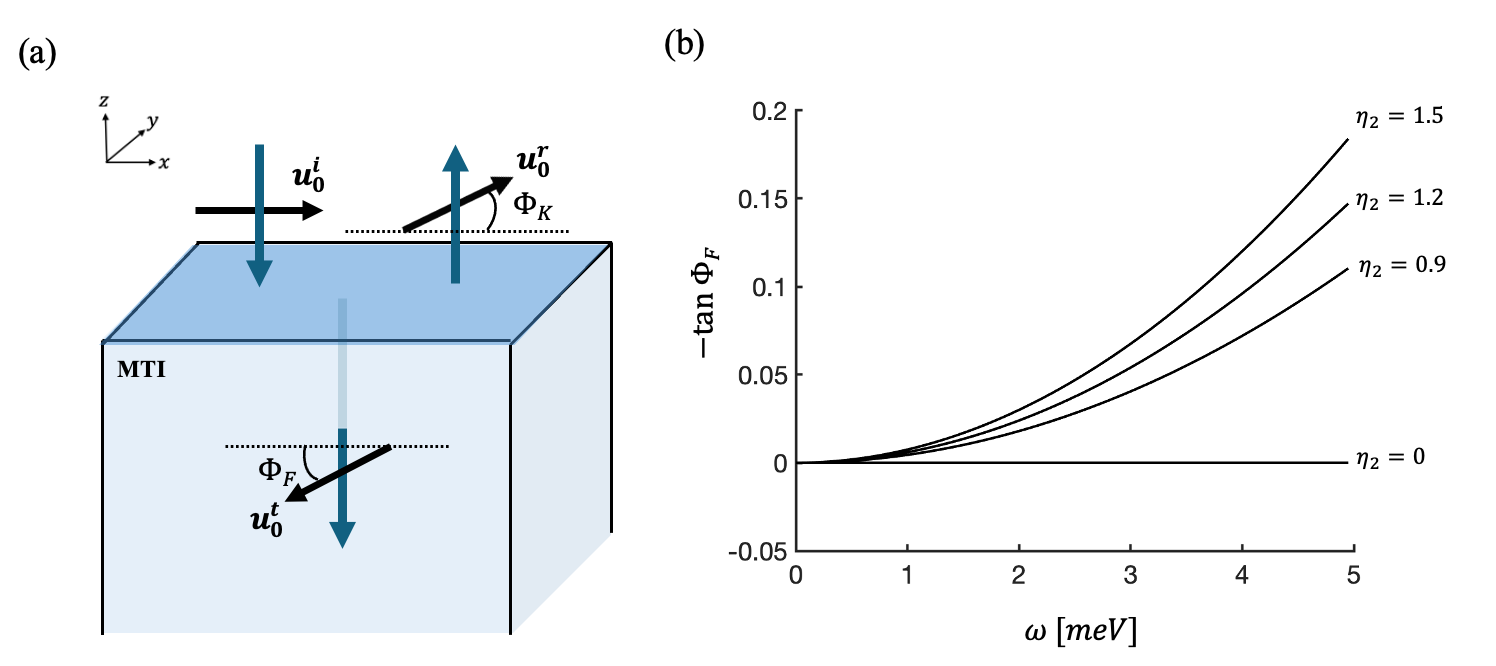}
 \caption{  (a) The setup of the acoustic Faraday effect due to the surface phonon Hall viscosity. The blue arrows depict the direction of propagation and the black arrows depict the polarization. (b) The acoustic Faraday angle $ \tan \Phi_F$ as a function of the injected elastic wave frequency $\omega$ for different $\eta_2$. $\eta_2$ is in units of $\frac{1}{\rho_0} \frac{1}{\hbar} \frac{1}{(\mu m)^2}
 $. 
 }
 \label{fig:Faraday}
 \end{figure*}

We numerically compute the transmitted elastic wave at $y=W$ using Eq.(\ref{eq:PsiyFinal}). We compute the angular momentum $L_y^t$ of ${\bf u}^t(W)$ near the interface, evaluated over a region spanning a few sites, from $N/2 - N_{\text{int}}$ to $N/2 + N_{\text{int}}$ with $ N_{\text{int}} = 7$ sites, e.g. $L_y^t$ is defined as
\begin{equation}
    L_y^t = \sum_{m=N/2 - N_{\text{int}}}^{N/2 + N_{\text{int}}} \Tilde{L}_y^t(z_m),
\end{equation}
where $\Tilde{L}_y^t(z_m)$ is given by
\begin{equation}
    \Tilde{L}_y^t(z_m) = 2 \hbar \text{Im} \Big[ u^*_z(z_m) u_x(z_m) \Big].
\end{equation}
The resulting phonon angular momentum $\Tilde{L}_y^t$ is shown in Fig. \ref{fig:CM2}(b) as a function of $z_m$ for an injected wave frequency of $2.5$ meV. We observe that $\Tilde{L}_y^t$ is localized near the interface $z/L \sim 0.5$ and is vanishingly small away from it. Further, the polarization changes from RCP to LCP once $\eta_0$ is flipped from positive to negative. We depict the interface mode contribution to $\Tilde{L}_y^t$ in green for $\eta_0 > 0$ and in red for $\eta_0 <0$, while the total contribution to $\Tilde{L}_y^t$ is shown in black for $\eta_0 > 0$ and in blue for $\eta_0 <0$. In Fig. \ref{fig:CM2}(c), we plot $L_y^t$, summed over $2 N_{\text{int}}$ sites near the interface, highlighted by the light blue shaded region in \ref{fig:CM2}(b). We find the total polarization arising from all phonon modes ($N_k$ modes), shown in black, is close to $\pm \hbar$. The dominant contribution to $L_y^t$ originates from the interface mode, plotted in green. Thus, the signal detected near the interface is primarily determined by the interface RCP mode for $\eta_0 > 0$, where the polarization of the interface mode is preferentially filtered. Reversing the interface magnetization ($\eta_0 <0$) instead filters the LCP mode. This behavior is evident in  Fig. \ref{fig:CM2}(c), where changing the sign of the surface PHV $\eta_0$ flips the polarization of the detected mode from RCP to LCP. We also observe that the phonon angular momentum $L_y$ vanishes at $\eta_0$, consistent with our previous argument.

\begin{figure*}
\includegraphics[width=\textwidth]{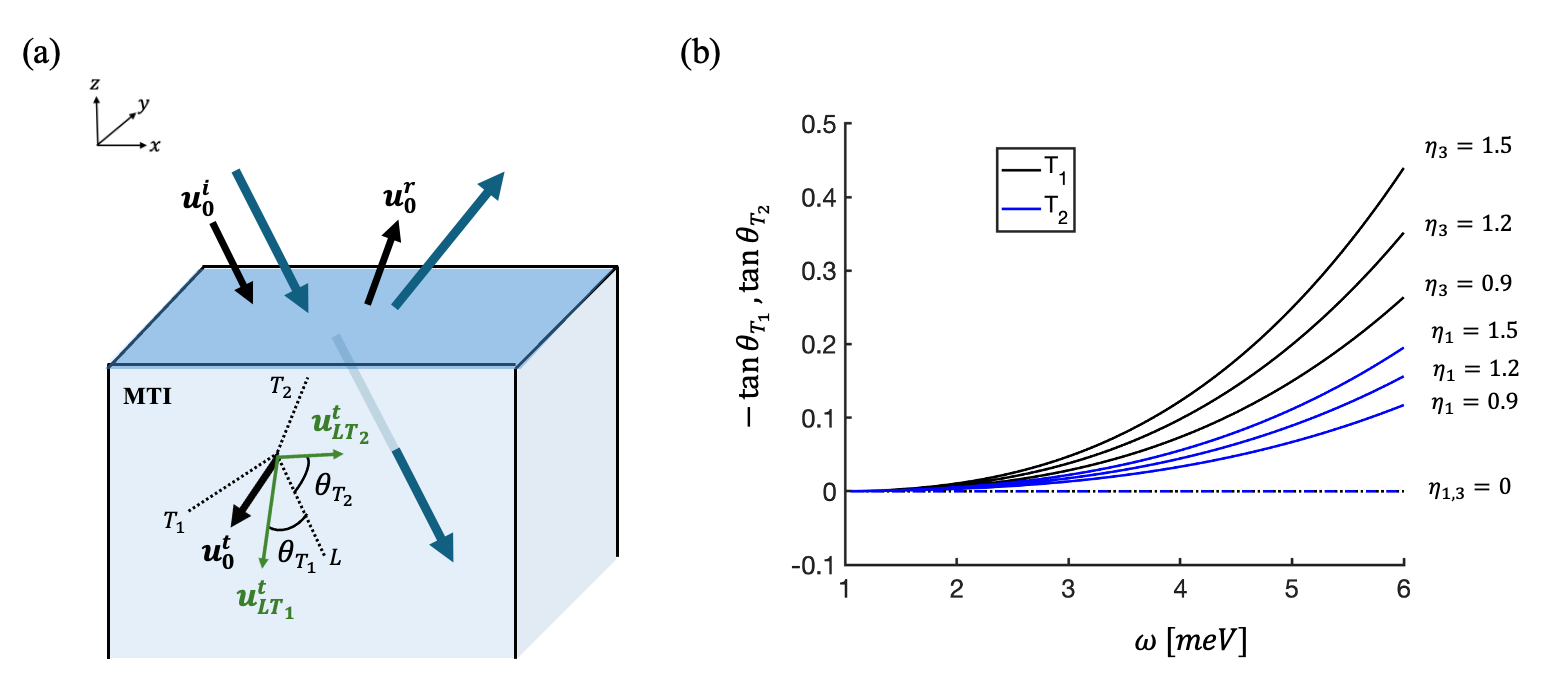}
 \caption{(a) The setup for the longitudinal-transverse mode conversion. The blue arrows depict the direction of propagation and the black arrows depict the polarization. The green arrows for $u_{LT_{1,2}}^t$ depicts the projection of the polarization $u_0^t$ on the $LT_{1,2}$ planes. The angles made by $u_{LT_{1,2}}$ with the $L$ axis are $\theta_{T_{1,2}}$ (b) The conversion efficiency characterized by angles $\theta_{T_1}, \theta_{T_2} $ along the two transverse directions as a function of $\omega$ for different $\eta_0$. We use $\eta_{1,3}$ in units of $\frac{1}{\rho_0} \frac{1}{\hbar} \frac{1}{(\mu m)^2}
 $ and $c_t = 3500$ m/s, $k_z = 0.03$ nm$^{-1}$, $\eta_2=0$. 
}
 \label{fig:Conversion}
 \end{figure*}

\section{Generalized Surface acoustic Faraday rotation} \label{sec:Faraday}

In this section, we solve the scattering problem at the surface of a magnetic TI sandwich structure \cite{yu2010quantized,chang2013experimental,zhang2012tailoring,morimoto2015,xiao2018realization,mogi2017magnetic,mogi2017tailoring,zhuo2023axion,jiang2020concurrence,kou2015magnetic} or MnBi$_2$Te$_4$ \cite{deng2020quantum,liu2020robust,otrokov2019prediction,zhang2019topological,li2019intrinsic,gong2019experimental,liu2020robust}. We consider the setup shown in Fig.\ref{fig:Schematic}(b) and (c) or Fig.\ref{fig:Faraday}(a), in which a linearly polarized transverse elastic wave, traveling along the $z$-direction is injected at the surface of a magnetic TI material ($z \leq 0$) from the trivial side ($z>0$). The displacement fields for the injected, reflected and transmitted waves are given by
\begin{equation}
    \textbf{u} = \begin{cases}
        \textbf{u}_0^i e^{i \left( -k z - \omega t\right) } + \textbf{u}_0^r e^{i \left( k z - \omega t\right) } &  z>0 \\
       \textbf{u}_0^t e^{i \left( -k z - \omega t\right) } &  z<0 .
    \end{cases} \label{eq_SM:AFEu}
\end{equation}
There are two boundary conditions at $z=0$: (1) Continuity of $\textbf{u}$ at $z=0$ is given by
\begin{equation}
    \textbf{u}_0^i + \textbf{u}_0^r = \textbf{u}_0^t ;\label{eq_SM:AFECont}
\end{equation}
(2) Since the surface effects are captured by Eq.(\ref{eq:HNY}), the second boundary condition is obtained by integrating the equation of motion given by 
Eqs.(\ref{eq:H0}-\ref{eq:HNY}) over a small region across the surface, i.e. 
\begin{equation}
    \int_{-\epsilon}^{+\epsilon} dz \left( H_0 +  2 i \omega H_{\text{PHV}}(\partial_z) \right) \textbf{u} = \omega^2 \int_{-\epsilon}^{+\epsilon} dz \textbf{u} , 
\end{equation}
which implies
\begin{align}
    &\int_{-\epsilon}^{+\epsilon} dz \Big[ -d \partial_z^2 u_x - 2 i \omega \eta_2 \delta(z) \partial_z^2 u_y   \Big] = \int_{-\epsilon}^{+\epsilon} dz \omega^2 u_x =0 \label{eq_SM:AFE1} \\
    &\int_{-\epsilon}^{+\epsilon} dz \Big[ -d \partial_z^2 u_y + 2 i \omega \eta_2 \delta(z) \partial_z^2 u_x   \Big] = \int_{-\epsilon}^{+\epsilon} dz \omega^2 u_y = 0.\label{eq_SM:AFE2} 
\end{align}
We note that Eqs. (\ref{eq_SM:AFE1},\ref{eq_SM:AFE2}) depend only on $\eta_2$ and are independent of $\eta_{1,3}$. This follows from Eq.(\ref{eq:HNY}), where $\eta_{1,3}$ always appear multiplied by $k_x$ or $k_y$. Since we consider pure normal incidence, $k_{x} = k_y = 0$ and consequently, Eqs. (\ref{eq_SM:AFE1},\ref{eq_SM:AFE2}) contain no dependence on $\eta_{1,3}$. We see that $u_z$ is decoupled from $u_{x,y}$ in the above equations, therefore we take $\textbf{u}_0 = (u_x, u_y)^T$ for the transverse wave. Substituting Eqs.(\ref{eq_SM:AFEu},\ref{eq_SM:AFECont}) into Eqs.(\ref{eq_SM:AFE1},\ref{eq_SM:AFE2}), we get for $\epsilon \rightarrow 0$
\begin{align}
& d k \left( 2 u_{0x}^i -  2 u_{0x}^t \right) + 2 \omega k^2 \eta_2 u_{0y}^t = 0 \label{eq_SM:AFE1a} \\
& d k \left( 2 u_{0y}^i -  2 u_{0y}^t \right) - 2 \omega k^2 \eta_2 u_{0x}^t = 0 , \label{eq_SM:AFE2a}
\end{align}
where we have used the continuity condition in Eq.(\ref{eq_SM:AFECont}). We can rewrite Eqs.(\ref{eq_SM:AFE1a},\ref{eq_SM:AFE2a}) in a matrix form as
\begin{equation}
    \begin{pmatrix}
        u_{0x}^t \\
        u_{0y}^t
    \end{pmatrix} = \frac{1}{1 + \left( \frac{\eta_2 k \omega}{d}\right)^2}\begin{pmatrix}
        1 & \frac{\eta_2 k \omega}{d} \\
        -\frac{\eta_2 k \omega}{d} & 1
    \end{pmatrix} \begin{pmatrix}
        u_{0x}^i \\
        u_{0y}^i
    \end{pmatrix} .\label{eq_SM:AFEMatrix}
\end{equation}
We take the initial elastic wave as $u_{0x}^i \neq 0, u_{0y}^i =0 $ and Eq.(\ref{eq_SM:AFEMatrix}) becomes
\begin{align}
    & u_{0x}^t = \frac{1}{1 + \left( \frac{\eta_2 k \omega}{d}\right)^2} u_{0x}^i \\
    & u_{0y}^t = -\frac{1}{1 + \left( \frac{\eta_2 k \omega}{d}\right)^2} \frac{\eta_2 k \omega}{d} u_{0x}^i .
\end{align}
Therefore, injecting an acoustic wave linearly polarized in the $x$-direction can lead to a rotation of the polarization in the $xy$-plane and the rotation angle of the polarization vector is determined by surface PHV as
\begin{equation}
    \tan \Phi_F = \frac{u_{0y}^t}{u_{0x}^t} = -  \eta_2 \frac{\omega^2}{d^{3/2}},
\end{equation}
where we have used the dispersion relation of the acoustic wave as $\omega = \sqrt{d} k$, which has been obtained by solving Eq.(\ref{eq:H0}) with $k_x = k_y = 0$ and $\partial_z \rightarrow i k$. In Fig.\ref{fig:Faraday}(b), we plot the Faraday angle as a function of $\omega$ for various $\eta_2$ values. The Faraday angle is linear in the PHV coefficient $\eta_2$ and is quadratic in the injected wave frequency $\omega$. As expected, when the surface PHV $\eta_2 = 0$, there is no acoustic Faraday rotation. Similarly, the acoustic Kerr angle becomes 
\begin{equation}
    \tan \Phi_K = \frac{u_{0y}^r}{u_{0x}^r} = +  \eta_2 \frac{\omega^2}{d^{3/2}}
\end{equation}
which is the opposite of the acoustic Faraday angle.

\section{Longitudinal-transverse mode conversion} \label{sec:Conversion}
In this section, we present the conversion between longitudinal and transverse modes for oblique incident acoustic waves at the surface of a magnetic TI sandwich structure \cite{yu2010quantized,chang2013experimental,zhang2012tailoring,morimoto2015,xiao2018realization,mogi2017magnetic,mogi2017tailoring,zhuo2023axion,jiang2020concurrence,kou2015magnetic} or MnBi$_2$Te$_4$ \cite{deng2020quantum,liu2020robust,otrokov2019prediction,zhang2019topological,li2019intrinsic,gong2019experimental,liu2020robust}. We consider the setup shown in Fig.\ref{fig:Conversion}(a), in which a linearly polarized longitudinal acoustic wave, traveling in the $xz$-plane is injected at the surface of a magnetic TI material. We consider the isotropic approximation of Eq.(\ref{eq:isoparameters}) in the bulk so that the longitudinal and transverse modes are decoupled in the bulk. We also choose $\eta_2 =0$ for simplicity. The longitudinal ($L$) and 2 transverse modes ($T_{1,2}$) are given by
\begin{align}
    &\textbf{u}_{0L} = \frac{1}{\sqrt{k^2 +k_z^2}}\begin{pmatrix}
        -k \\ 0 \\ k_z
    \end{pmatrix}, \\ &\textbf{u}_{0T_1} = \frac{1}{\sqrt{k^2 +k_z^2}}\begin{pmatrix}
        k_z \\ 0 \\ k
    \end{pmatrix}, \nonumber \\
    &\textbf{u}_{0T_2} = \begin{pmatrix}
        0 \\ 1 \\ 0
    \end{pmatrix}, 
\end{align}
and are decoupled in the bulk. The basis formed by the longitudinal ($L$) and two transverse modes ($T_1,T_2$) is depicted in Fig. \ref{fig:Conversion}(a) as black dotted lines. In this basis, Eqs.(\ref{eq:H0}-\ref{eq:HNY}) become
\begin{equation}
    H_0  = \begin{pmatrix} 
     c_l^2 \left( k^2 -  \partial_z^2 \right) & 0  &  0 \\
    0  & c_t^2 \left( k^2 -  \partial_z^2 \right) & 0  \\
     0 & 0 & c_t^2 \left( k^2 -  \partial_z^2 \right) 
    \end{pmatrix},\label{eq:H0CM}
\end{equation}
and 
\begin{equation}
    H_{\text{PHV}}  =  \delta(z) \begin{pmatrix} 
     0  &  \eta_3 k^2 & -\eta_1 \frac{k^3}{\sqrt{k^2+kz^2}}\\ 
       -\eta_3 k^2 & 0 & \eta_1 \frac{k^2 k_z}{\sqrt{k^2+kz^2}} \\
     \eta_1 \frac{k^3}{\sqrt{k^2+k_z^2}}  & -\eta_1 \frac{k^2 k_z}{\sqrt{k^2+k_z^2}} & 0
    \end{pmatrix}. \label{eq:HNYCM}
\end{equation}
The acoustic waves at the trivial side ($z>0$) and the magnetic TI side ($z<0$) are given by
\begin{equation}
    \textbf{u} = \begin{cases}
        \textbf{u}_0^i e^{-i k_z z}  e^{i \left( k x - \omega t\right) } + \textbf{u}_0^r e^{i k_z z} e^{i \left( k x - \omega t\right) } &  z>0 \\
       \textbf{u}_0^t e^{i k_z z} e^{i \left( k x - \omega t\right) } &  z<0 .
    \end{cases} \label{eq_SM:CMu}
\end{equation}

There are two boundary conditions at $z=0$: (1) Continuity of $\textbf{u}$ at $z=0$ is given by
\begin{equation}
    \textbf{u}_0^i + \textbf{u}_0^r = \textbf{u}_0^t ;\label{eq_SM:CMCont}
\end{equation}
(2) The integration of the equation of motion given by 
Eqs.(\ref{eq:H0CM}-\ref{eq:HNYCM}) over a small region across the surface gives
\begin{equation}
    \int_{-\epsilon}^{+\epsilon} dz \left( H_0 +  2 i \omega H_{\text{PHV}}\right) \textbf{u} = \omega^2 \int_{-\epsilon}^{+\epsilon} dz \textbf{u}, 
\end{equation}
with $\textbf{u} = (u_L, u_{T_1},u_{T_2})^T$, which implies

\begin{widetext}
\begin{align}
    &\int_{-\epsilon}^{+\epsilon} dz \Bigg[ -c_l^2 \partial_z^2 u_L + 2 i \omega \eta_3 \delta(z) k^2 u_{T_1}  - 2 i \omega \eta_1 \delta(z) \frac{k^3}{\sqrt{k^2+k_z^2}} u_{T_2} \Bigg] = \int_{-\epsilon}^{+\epsilon} dz \omega^2 u_L   =0 \label{eq_SM:CM1} \\
    &\int_{-\epsilon}^{+\epsilon} dz \Bigg[-2 i \omega \eta_3 k^2 \delta(z) u_L  - c_t^2 \partial_z^2 u_{T_1} + 2 i \omega \eta_1  \delta(z) \frac{k^2 k_z}{\sqrt{k^2+k_z^2}} u_{T_2}   \Bigg] = \int_{-\epsilon}^{+\epsilon} dz \omega^2 u_{T_1} = 0 \label{eq_SM:CM2} \\
    &\int_{-\epsilon}^{+\epsilon} dz \Bigg[ 2 i \omega \eta_1 \frac{k^3}{\sqrt{k^2+k_z^2}} \delta(z) u_L - 2 i \omega \eta_1 \delta(z) \frac{k^2 k_z}{\sqrt{k^2+k_z^2}} u_{T_1} - c_t^2 \partial_z^2 u_{T_2} \Bigg] = \int_{-\epsilon}^{+\epsilon} dz \omega^2 u_{T_2} = 0. \label{eq_SM:CM3} 
\end{align}
Eqs.(\ref{eq_SM:CM1},\ref{eq_SM:CM2},\ref{eq_SM:CM3}) can be evaluated as
\begin{align}
       -c_l^2 \partial_z u_L |_{-\epsilon}^{+\epsilon} + 2 i \omega \eta_3 k^2 u_{T_1} |_{z=0}   - 2 i \omega \eta_1 \frac{k^3}{\sqrt{k^2+k_z^2}}  u_{T_2} |_{z=0} & =0 \label{eq_SM:CM1a} \\
     -2 i \omega \eta_3 k^2  u_L|_{z=0}  - c_t^2 \partial_z u_{T_1} |_{-\epsilon}^{+\epsilon} + 2 i \omega \eta_1 \frac{k^2 k_z}{\sqrt{k^2+k_z^2}} u_{T_2} |_{z=0}  &= 0 \label{eq_SM:CM2a} \\
     2 i \omega \eta_1 \frac{k^3}{\sqrt{k^2+k_z^2}}  u_L |_{z=0} - 2 i \omega \eta_1 i \frac{k^2 k_z}{\sqrt{k^2+k_z^2}} u_{T_1}|_{z=0} -c_t^2 \partial_z u_{T_2} |_{-\epsilon}^{+\epsilon}  &= 0. \label{eq_SM:CM3a}
\end{align}

Substituting Eqs.(\ref{eq_SM:CMu},\ref{eq_SM:CMCont}) into Eqs.(\ref{eq_SM:CM1a},\ref{eq_SM:CM2a},\ref{eq_SM:CM3a}), we get for $\epsilon \rightarrow 0$
\begin{align}
 u_{0L}^i &= u_{0L}^t - \frac{\eta_3 k^2}{c_l^2 k_z} u_{0T_1}^t + \frac{\eta_1 k^3}{c_l^2 k_z \sqrt{k^2+k_z^2}}  u_{0T_2}^t \label{eq_SM:CM1b} \\
 u_{0T_1}^i &= \frac{\eta_3 k^2}{c_t^2 k_z} u_{0L}^t + u_{0T_1}^t - \frac{\eta_1 k^2}{c_t^2 \sqrt{k^2+k_z^2}}  u_{0T_2}^t  \label{eq_SM:CM2b} \\
 u_{0T_2}^i &= - \frac{\eta_1 k^3}{\sqrt{k^2+k_z^2}}u_{0L}^t + \frac{\eta_1 k^2}{c_t^2} u_{0T_1}^t +  u_{0T_2}^t . \label{eq_SM:CM3b}
\end{align}
We can rewrite Eqs.(\ref{eq_SM:CM1b},\ref{eq_SM:CM2b},\ref{eq_SM:CM3b}) to linear order in $\eta_{1,3}$ as
\begin{equation}
    \begin{pmatrix}
        u_{0L}^t \\
        u_{0T_1}^t \\
        u_{0T_2}^t
    \end{pmatrix} = \begin{pmatrix}
        1 &  \frac{  \eta_3 \omega k^2 }{c_l^2 k_z} & -\frac{  \eta_1 \omega k^3 }{c_l^2 k_z \sqrt{k^2+k_z^2}} \\
         -\frac{  \eta_3 \omega k^2 }{c_t^2 k_z} & 1 &  \frac{  \eta_1 \omega k^2 }{c_t^2 \sqrt{k^2 + k_z^2}} \\  \frac{  \eta_1 \omega k^3 }{c_t^2 k_z \sqrt{k^2+k_z^2}} & -\frac{  \eta_1 \omega k^2 }{c_t^2 \sqrt{k^2+k_z^2}} & 1 
    \end{pmatrix} \begin{pmatrix}
        u_{0L}^i \\
        u_{0T_1}^i \\
        u_{0T_2}^i
    \end{pmatrix}. \label{eq_SM:CMMatrix}
\end{equation}
\end{widetext}

We take the initial acoustic wave to be $\textbf{u}_0^i = \textbf{u}_{0L}$ and find the transmitted wave is given by
\begin{align} \label{eq_SM:TransmittedCM}
    u_{0L}^t &=  u_{0L}^i\nonumber \\
    u_{0T_1}^t &= -\frac{\eta_3 \omega k^2}{c_t^2 k_z} u_{0L}^i = -\eta_3 \frac{\omega}{c_t^2 k_z} \left( \omega^2/c_t^2 - k_z^2 \right) u_{0L}^i \nonumber \\
    u_{0T_2}^t &=  \frac{\eta_1 \omega k^3}{c_t^2 k_z \sqrt{k^2+k_z^2}} u_{0L}^i = \eta_1 \frac{1}{c_t k_z} \left( \omega^2/c_t^2 - k_z^2\right)^{3/2} u_{0L}^i.
\end{align}
We find from Eq.(\ref{eq_SM:TransmittedCM}) that when a longitudinal mode is injected, the surface PHV leads to the generation of transverse modes, a special case of which was presented in Ref.\cite{shapourian2015viscoelastic} that corresponds to the case with $\eta_1 = \eta_2, \eta_3 = 0$ in our work. We define the angles $\theta_{T_1}, \theta_{T_2}$ as
\begin{align}
    &\tan \theta_{T_1} = \frac{u_{0T_1}^t}{u_{0L}^t} = -\eta_3 \frac{\omega}{c_t^2 k_z} \left( \omega^2/c_t^2 - k_z^2 \right) \\
    &\tan \theta_{T_2} = \frac{u_{0T_2}^t}{u_{0L}^t} = \eta_1 \frac{1}{c_t k_z} \left( \omega^2/c_t^2 - k_z^2\right)^{3/2},
\end{align}
which characterizes the conversion efficiency in two transverse directions. As shown in Fig. \ref{fig:Conversion}(a), the transmitted acoustic wave ${\bf u_0^t}$ can be projected onto the $L T_{1}$ and $L T_{2}$ planes and the projected modes are $u_{LT_1}^t$ and $u_{LT_2}^t$, respectively. The projected modes $u_{LT_{1,2}}^t$ make angles $\theta_{T_{1,2}}$ with the longitudinal axis $L$. In Fig. \ref{fig:Conversion}(b), we plot $\theta_{T_1}, \theta_{T_2}$ as a function of $\omega$ for various $\eta_{1,3}$ values. Similar to the acoustic Faraday effect, the longitudinal-transverse conversion efficiency is linear in the surface PHV and is quadratic in the injected elastic wave frequency $\omega$. In the absence of the surface PHV $\eta_{1,3}$, there is no longitudinal to transverse mode conversion.

\section{Discussion and Conclusion}\label{sec:Conclusion}
In this work, we theoretically identified a previously unexplored phonon polarization-filter mechanism using interface phonon modes in magnetic TIs. We demonstrated that the PHV at the surface or interface, intrinsic to these materials, induces an interface phonon mode, with its frequency lower than the bulk mode frequency. This phonon mode is localized at the interface and carries phonon angular momentum pertaining to a specific circularly polarization. Consequently, magnetic TIs naturally serve as phonon polarization-filters, controlling phonon polarization through surface magnetic orders. Experimentally, the predicted interface phonon modes can be probed using established ultrasound techniques \cite{boyd1966attenuation,boiteux1971acoustical,lee1999discovery}, surface acoustic wave platforms \cite{datta1986surface,morgan2010surface}, or time-resolved pump-probe measurements sensitive to phonon polarization \cite{thomsen1984coherent,forst2011nonlinear}. More broadly, our results establish surface PHV as a powerful route for controlling phonon polarization, opening pathways toward intrinsic phononic devices such as waveplates, polarization-selective elements, and axial and chiral phonon sources based on magnetic TI materials. 

We also generalized the acoustic Faraday rotation and the longitudinal-transverse mode conversion in magnetic TI materials to incorporate all the symmetry-allowed terms. Since the Faraday angle is proportional to the surface PHV given by the surface magnetization, the acoustic Faraday effect in magnetic TI sandwich structures and MnBi$_2$Te$_4$ thin films is expected to cancel when the top and bottom surface magnetizations are antiparallel and would add up when they are parallel, producing a finite rotation of the incident polarization vector. This behavior is directly analogous to the even-odd effects in the Faraday rotation response from axion electrodynamics \cite{chen2024even,zhao2021even,li2024fabrication,yang2024intrinsic,mei2024electrically,lin2022direct,ovchinnikov2021intertwined}. We further presented a generalized mode conversion for realistic materials, where the ability of each mode to convert into the other can significantly expand the range of practical application in phononics.

\section{Acknowledgments}
We thank M.K. Joshi, C.Z. Chang, X.L. Qi, H. Nieh and B.H. Yan for helpful discussions. A.C. and C.-X.L. also acknowledge support from NSF grant via the grant number DMR-2241327 and the ONR Award (N000142412133). 

\bibliographystyle{apsrev4-2}
\bibliography{ref}

\end{document}